\documentclass[doublespacing]{elsart}
\usepackage[english]{babel}

\usepackage{epsfig}
\usepackage{graphics}

\usepackage{amssymb}

\begin{document}

\begin{frontmatter}

\title{High Power Microwave and Optical Volume Free Electron Lasers (VFELs)}

\author{V.G. Baryshevsky}

\address{Research Institute for Nuclear Problems, Belarusian State
University, 11~Bobruiskaya Str., Minsk 220030, Belarus}
\ead{bar@inp.bsu.by, v\_baryshevsky@yahoo.com}

%\affiliation{Research Institute for Nuclear Problems....}

%
\begin{abstract}
The use of a non-one-dimensional distributed feedback, arising
through Bragg diffraction in spatially periodic systems (natural
and artificial (electromagnetic, photonic) crystals), forms the
foundation for  the development of volume free electron lasers
(VFELs). The present review  addresses the basic principles of
VFEL theory and describes the promising potential of VFELs as the
basis for the development of high-power microwave and optical
sources.
\end{abstract}
\tableofcontents
\end{frontmatter}

\section{Introduction}

\begin{figure}[htbp]
\begin{center}
\includegraphics[scale=1.]{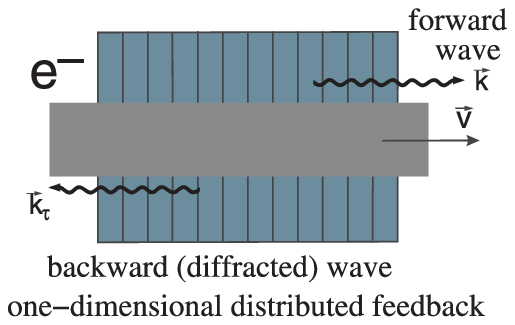}\\%qquad
\vspace{0.5cm}\includegraphics[scale=0.3]{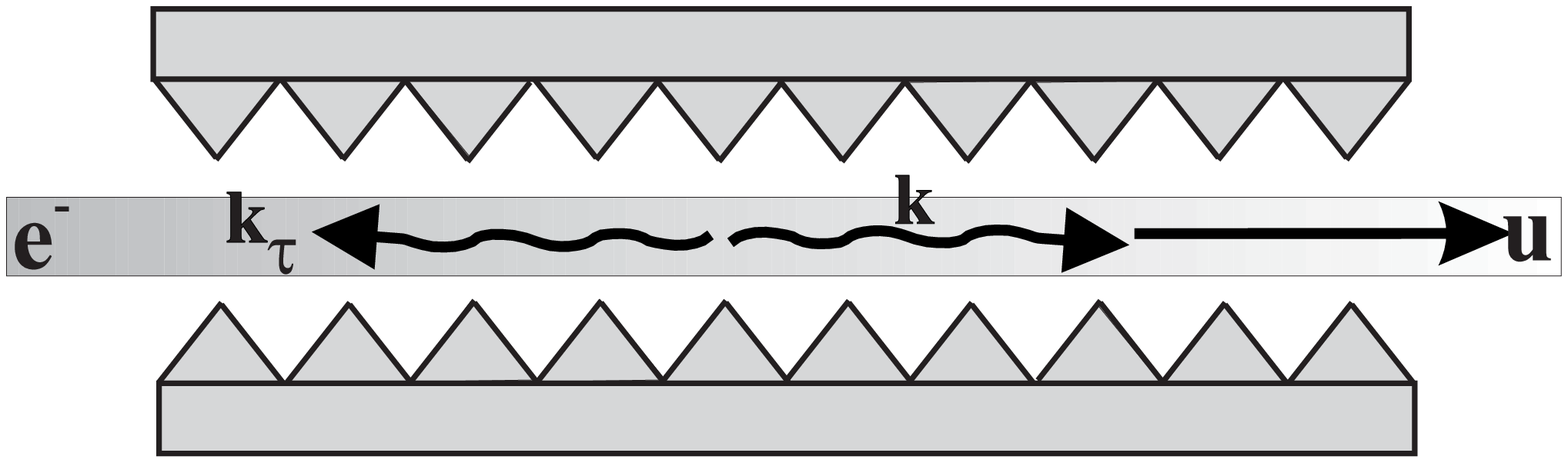} \caption{}
\label{rotation1}
\end{center}
\end{figure}

Research and development of microwave generators using radiation
from an electron beam in a periodic slow--wave circuit
(traveling--wave tube, backward wave oscillator, etc.) has a long
history (see e.g. \cite{2_bvg,gold,ben}).
One of the characteristic features of such generators is the use
of distributed feedback (DFB) where the electromagnetic waves
produced by the electron beam are collinear with the direction of
the beam motion (one-dimensional distributed feedback; see Fig.
\ref{rotation1}).

It is well known that every radiating system is defined by its
eigenmodes and by the so-called dispersion equation, which
describes the relation between the frequency $\omega$ and the wave
vector $\vec k$ ($\omega=\omega(\vec k)$ or $\vec k=\vec
k(\omega)$).
A thorough analysis of the dispersion equation shows that:
\begin{enumerate}
\item
dispersion equation for a FEL \cite{laser_1} in the Compton
regime coincides with that for a conventional traveling wave tube
amplifier (TWTA);
\item
 FEL gain (increment of electron beam
instability) in the Compton regime is proportional to $n_0
^{1/3}$, where $n_0$ is the electron beam density.
\end{enumerate}

It was first shown  in \cite{PLetters,Vacuum}  that the law of
electron beam instability can change significantly under the
conditions of a non-one-dimensional distributed feedback formed in
a two-- or three--dimensional periodic resonator (natural or
artificial (electromagnetic, photonic) crystal); see Fig.
\ref{volume}.

\begin{figure}[htbp]
\begin{center}
\includegraphics[scale=1.]{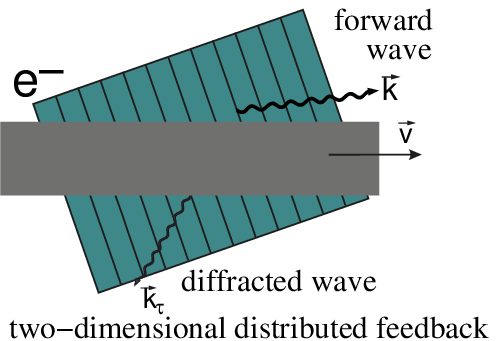}\\%qquad
\vspace{0.5cm}\includegraphics[scale=0.3]{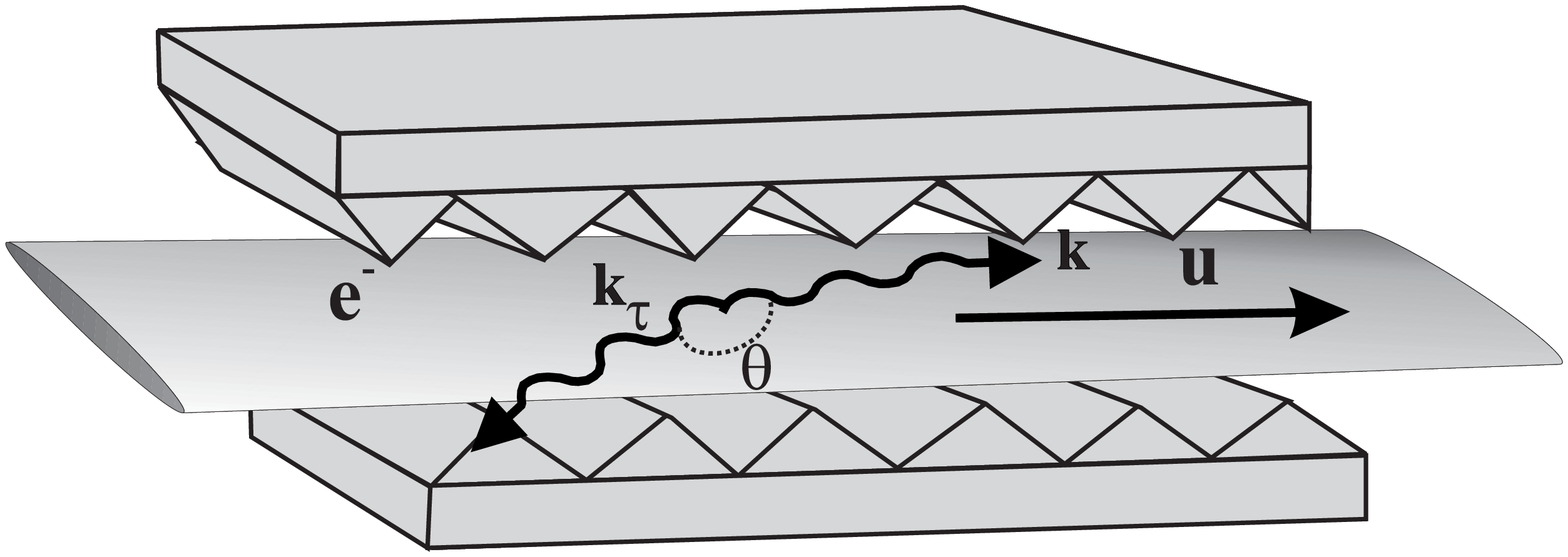}%\\
\end{center}
\end{figure}
\begin{figure}[htbp]
\begin{center}
\epsfxsize = 10 cm \centerline{\epsfbox{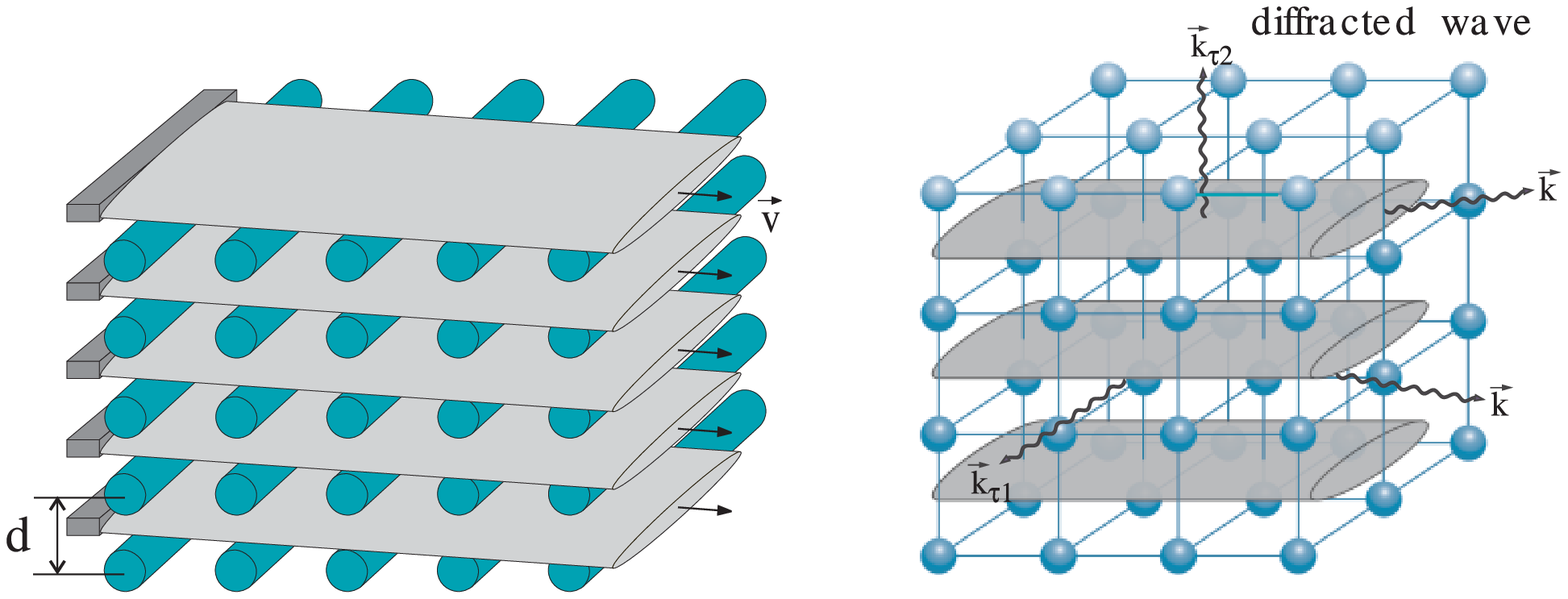}} \caption{}
\label{volume}
\end{center}
\end{figure}
\begin{figure}[h]
\begin{center}
%---\epsfxsize = 13 cm \
%---centerline{\epsfbox{multiwave.eps}}
\includegraphics[scale=0.6]{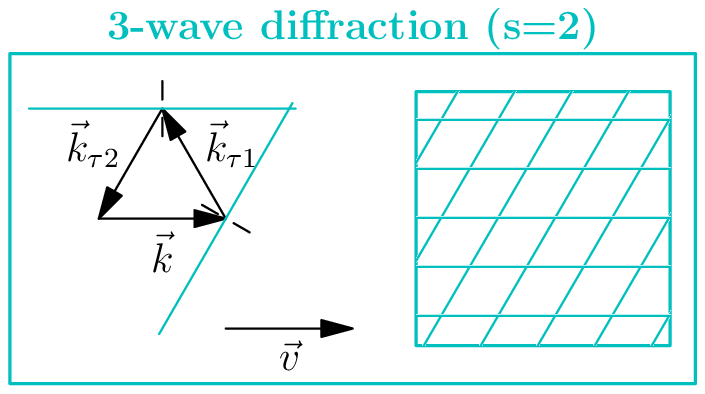}
\quad \includegraphics[scale=0.6]{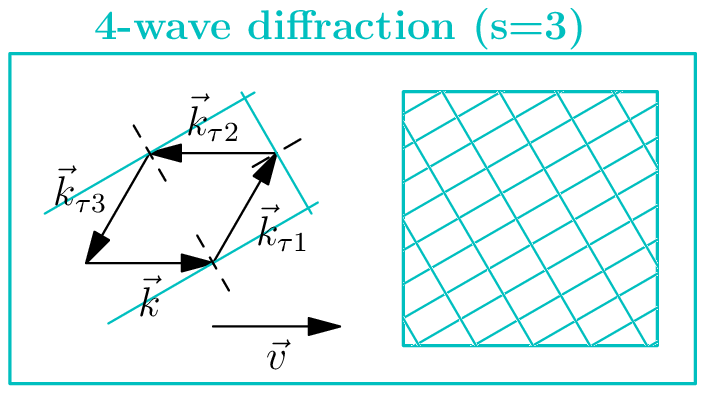}
\quad\includegraphics[scale=0.6]{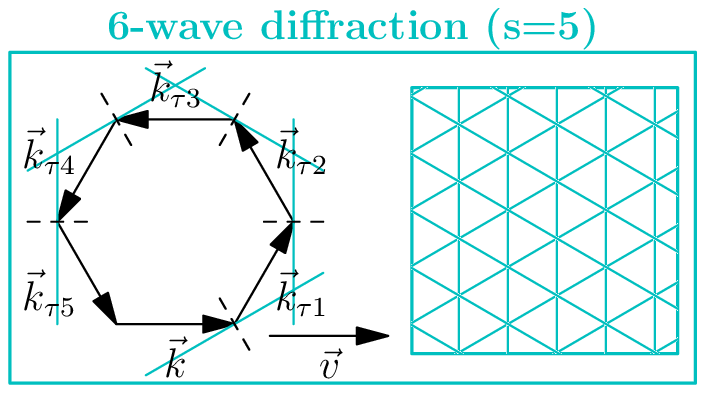}
\end{center}
\caption{} \label{multiwave}
\end{figure}

 It was shown that the gain
of electromagnetic waves changes sharply in the vicinity of the
points where the roots of the dispersion equation coincide. In
particular, according to \cite{PLetters,Vacuum}, in the case when
the DFB in a spatially periodic resonator is formed by the two
waves participating in the Bragg diffraction, the increment of
instability is proportional to $n_0^{1/4}$. If the DFB is formed
by $N$ number of waves (as a result of Bragg diffraction $s=N-1$
number of extra waves are produced (see Fig. \ref{multiwave})),
 then the increment of instability appears to be proportional to
$n_0^{\frac{1}{3+s}}$, provided that the roots of the dispersion
equation coincide (e.g., for two--wave Bragg diffraction, $N=2$
and $s=1$, for three--wave diffraction, $N=3$ and $s=2$ ).
This result is also valid for electron beams moving in  vacuum
near the surface of a spatially periodic medium, in a vacuum slot
made inside a periodic medium, or  between the diffraction
gratings of the resonator. The explicit expressions for the
starting current $j$ were obtained for the conditions of a
non-one-dimensional DFB, and it was shown that the threshold
currents in this case can be sharply reduced.

The advantages of VFEL are pronounced in a wide spectral range --
from microwaves to X-rays
\cite{PLetters,Vacuum,LANL98,bar1,bar4,laser_8}.

Providing the capabilities for frequency tuning,  using wide
electron beams (several e-beams), and reducing the threshold
current density, required to initiate lasing, VFELs offer a more
promising potential for  the development of more compact,
high-power and tunable radiation sources than conventional
electron vacuum devices.

In particular, experimental investigation of the properties of
electromagnetic crystals formed by periodically spaced dielectric
threads (wires) demonstrates that the quality factor of such
structures can be as high as $\sim 10^6 \div 10^8$ \cite{bar4}.
First experiments have been performed on exciting generation  in
different types of periodic structures (resonators made of
diffraction gratings with two different periods, resonators based
on a photonic crystal formed by metallic wires and foils) (for
details, see \cite{LANL98}).

Let us note that in  \cite{gin,gin2,gin3}, published some years
after \cite{PLetters,Vacuum}, the authors, considering  a
particular case of using a two-dimensional DFB for the
synchronization of radiation across a wide sheet electron beam,
 also came to the conclusion that such distributed feedback can
be used for developing microwave generators.

 Benefits given by VFELs:
\begin{enumerate}
\item
 volume FELs provide  frequency tuning by rotation of
the diffraction grating;
\item
 use of a non-one-dimensional DFB  reduces the generation
threshold and the size of the generation zone.
It appeared that if $N=1+s$ number of  waves participate in the
formation of a non-one-dimensional DFB, then there are $N$ number
of connected  waves propagating in the electromagnetic (photonic)
crystal (see Fig.\ref{multiwave} ).
This set of $N $ number of connected waves has $N$ number of
stationary states characterized by wave numbers $k_{i}(\omega)$,
($i=1, ...N$). If the relationship $(k_i-k_j)L=2 \pi m_{ij}$
($m_{ij}$ are the integers, which in the general case are not the
same) between these waves holds, then a significant change in the
threshold current value is possible. Let us note that in the case
of a one-dimensional DFD (see Fig. \ref{rotation1}), these
conditions transform into the requirement for the efficiency of a
one-dimensional DFB \cite{kog};
 \item VFELs allow the use of a
wide electron beam (or several beams) and diffraction gratings of
large volumes.
 Two- or three-dimensional diffraction gratings (artificial
crystals, now often called the electromagnetic or photonic
crystals) allow  distributing  the interaction over a  large
volume and overcoming the power restrictions on resonators (see
Fig. 2);
 \item
 VFELs can simultaneously generate radiation at several
frequencies;
 \item VFELs enable  effective mode selection in
oversized systems, where the radiation wave length is
significantly smaller than the resonator dimensions;
 \item
 use of electromagnetic (photonic) crystals with a spatially variable
 period allows increasing the efficiency of lasing  \cite{LANL98,FEL06}.
\end{enumerate}

\section{Radiative instability of beams moving in a spatially
periodic non-one-dimensional  resonator (two- or three-dimensional
electromagnetic (photonic) crystal)}

Let a relativistic  electron  beam of velocity $\vec{u}$ enter the
resonator having a form of a photonic crystal  with length $L$
(the $z$-axis is perpendicular to the crystal surface). Let this
photonic crystal have infinite transverse dimensions (crystal
plate) and lie in the interval ($0<z<L$).
The set of equations describing the interaction of electromagnetic
waves with the "crystal-beam" system consists of Maxwell's
equations and the equations of particle motion in the
electromagnetic field.
The dielectric susceptibility of a crystal has the form
$\varepsilon(\vec{r};\omega)=\sum_{\vec{\tau}}\varepsilon_{\tau}(\omega)\exp
(-i\vec{\tau}\vec{r})$, where $\vec{\tau}$ is the reciprocal
lattice vector, $\varepsilon_{\tau}$ is the Fourier expansion
coefficient of $\varepsilon(\vec r, \omega)$ and $\varepsilon(\vec
r, \omega)=1+\chi(\vec r, \omega)$, where $\chi (\vec r, \omega)$
is the space-periodic crystal polarizability, $\chi (\vec r,
\omega)=\sum_{\tau}\, \chi_{\tau}e^{-i\vec \tau\,\vec r}$. We
assume here for simplicity that $\varepsilon(\vec r, \omega)$ is
the scalar function. For concretness, let us further assume that
radiation is excited by quasi-Cherenkov (diffraction) radiation
(for details, see \cite{LANL98,bar1}). We also let
$|\chi_{\tau}\ll 1|$. Perturbations of the current and charge
densities in the linear field approximation may be written in the
form:
\begin{eqnarray}
\label{phys_1.1} & &
\delta\vec{j}(\vec{k};\omega)=e\sum\limits_{\alpha}\exp(-i\vec{k}\vec{r}_{\alpha
0} ) \left\{\delta \vec{v}_{\alpha}(\omega-\vec{k}\vec{u})
-i\vec{u}[\vec{k}\delta\vec{r}_{\alpha}(\omega-\vec{k}\vec{u})]\right\}\delta n(\vec{k};\omega)\nonumber\\
& & =e\sum\limits_{\alpha}\exp(-i\vec{k}\delta\vec{r}_{\alpha
0})\left\{-i[\vec{k}\vec{r}_{\alpha}(\omega-\vec{k}\vec{u})]\right\}.
\end{eqnarray}
Here $\delta\vec{j}(\vec{k};\omega)$ and $\delta
n(\vec{k};\omega)$ are the Fourier transformations of the
expressions
\[
\vec{j}(\vec{r};t)=e\sum\limits_{\alpha}\vec{v}_{\alpha}(t)\delta[\vec{r}-\vec{r}_{\alpha}(t)]\qquad
\mbox{and}\qquad
n(\vec{r};t)=\sum\limits_{\alpha}\delta[\vec{r}-\vec{r}_{\alpha}(t)]\,;
\]
$\vec{u}$ is the unperturbed electron (positron) velocity;
$\delta\vec{v}_{\alpha}$ and  $\delta\vec{r}_{\alpha}$ are
perturbations of the velocity and radius vectors, respectively,
arising due to the interaction with the radiation field:
\[
\vec{v}_{\alpha}(t)=\vec{u}+\delta\vec{v}_{\alpha}(t)\qquad
\vec{r}_{\alpha}(t)=\vec{r}_{\alpha
0}+\vec{u}t+\delta\vec{r}_{\alpha}(t).
\]
The subscript $\alpha$
denotes the number of the particle.

Using the equation of particle motion and formula (\ref{phys_1.1})
for current density, one can obtain a set of Maxwell's equations
that describes the interaction of electromagnetic waves and
particles in crystals:
\begin{eqnarray}
\label{phys_1.2} &
&k^2_{\tau}\vec{E}(\vec{k}_{\tau^{\prime}},\omega)-\vec{k}_{\tau^{\prime}}
[\vec{k}_{\tau^{\prime}}\vec{E}(\vec{k}_{\tau^{\prime}},\omega)]
-\frac{\omega^2}{c^2}\sum\limits_{\tau}
\varepsilon_{\tau}(\vec{k}_{\tau^{\prime}},\vec{\omega})\vec{E}(\vec{k}_{\tau+\tau^{\prime}},\omega)\nonumber\\
& &=-\frac{\omega^2_L}{\gamma
c^2}\vec{E}(\vec{k}_{\tau^{\prime}},\omega)-\left(\frac{\omega^2_L\vec{k}_{\tau^{\prime}}}
{\gamma c^2(\omega-\vec{k}_{\tau^{\prime}}
\vec{u})}+\frac{\omega^2_L(\vec{k}_{\tau^{\prime}}C^2-\omega^2)}
{\gamma C^4(\omega-\vec{k}_{\tau^{\prime}} \vec{u})^2}\right)\nonumber\\
& &\times\left[(\vec{u}\vec{E}(\vec{k}_{\tau^{\prime}},\omega))-
\left(\frac{\omega^2_L\vec{u}}{\gamma
c^2(\omega-\vec{k}_{\tau^{\prime}}
\vec{u})}\right)(\vec{k}_{\tau^{\prime}}\vec{E}(\vec{k}_{\tau^{\prime}},\omega))\right],\nonumber\\
& &\vec{\tau}\,^{\prime}  =  0,
\vec{\tau}_1,\vec{\tau}_2,...,\qquad \vec k_{\tau}=\vec
k+\vec\tau\,.
\end{eqnarray}

The set of equations (\ref{phys_1.2}) describes a situation when
the DFB is formed by many diffracted waves (note that in the
absence of an electron beam, the case of multi-wave diffraction
was studied in detail; see e.g. \cite{132}). However, the analysis
of such a general situation is very complicated, and so here we
consider only a two-wave distributed feedback. This allows us to
obtain all the main characteristics of VFELs analytically and show
the advantages of non-one-dimensional geometry of distributed
feedback over the one-dimensional one.

 So let us consider specifically the generation
of a $\sigma$-polarized wave (the wave with the polarization
vector  orthogonal to the diffraction plane, i.e., the plane where
vectors $\vec k$ and $\vec\tau$ lie) for the geometry of the
so-called two-beam Bragg diffraction \cite{L_23}, where two strong
waves are excited, and diffraction occurs by the set of
crystallographic planes, determined by the reciprocal lattice
vector~~$\vec\tau$ (see Figs \ref{fig.laser_1},
\ref{fig.laser_2}). In this case, the set of Maxwell's equations,
describing two-wave diffraction in crystals traversed by the beam,
can be written as
\begin{eqnarray}
\label{phys_1.3} &
&\left(k^2c^2-\omega^2\varepsilon_0+\frac{\omega^2_L}{\gamma}+\frac{\omega^2_L}{\gamma}
\frac{(\vec{u}\vec{e}\,^{\prime}_{\sigma})^2}{c^2}\frac{k^2c^2-\omega^2}{(\omega-\vec{k}\vec{u})^2}\right)E_{\sigma}-
\omega^2\varepsilon_{\tau}E_{\sigma}^{\tau}=0\nonumber\\
& &-\omega^2\varepsilon_{-\tau}E_{\sigma}+\left(K^2_{\tau}
c^2-\omega^2\varepsilon_0+\frac{\omega^2_L}{\gamma}+
\frac{\omega^2_L}{\gamma}\frac{(\vec{u}\vec{e}_{\sigma})^2}{e^2}
\frac{k^2_{\tau}c^2-\omega^2}{(\omega-\vec{k}_{\tau}\vec{u})^2}\right)\vec{E}_{\sigma}^{\tau}=0.\nonumber\\
\end{eqnarray}
In equation (\ref{phys_1.3}),
$E_{\sigma}=\vec{E}(\vec{k},\omega)\cdot\vec{e}_{\sigma}$,
$E_{\sigma}^{\tau}=\vec{E}(\vec{k}+\vec{\tau},\omega)\cdot\vec{e}_{\sigma}$,
$ \vec{e}_{\sigma}\parallel[\vec{k}\vec{\tau}]$, and
$\omega^2_L=4\pi e^2n_0/m$, where $n_0$ is the average electron
density in the beam. According to (\ref{phys_1.3}), the system
"crystal-particle beam" may be considered as an active medium with
dielectric susceptibility
\begin{eqnarray*}
\tilde{\varepsilon}_0(\vec{k},\omega)-1=\varepsilon_0-1-\frac{\omega^2_L}{\gamma\omega^2}-\frac{\omega^2-L}{\gamma\omega^2}
\frac{(\vec{u}\vec{e}_{\sigma})^2}{c^2}\frac{k^2c^2-\omega^2}{(\omega-\vec{k}\vec{u})^2},~ \tilde{\varepsilon}_{\tilde{\tau}}
=\varepsilon_{\tau}=\chi_{\tau}\\
\tilde{\varepsilon}_0(\vec{k}_{\tau},\omega)-1
=\varepsilon_0-1-\frac{\omega^2_L}{\gamma\omega^2}-\frac{\omega^2_L}{\gamma\omega^2}\frac{(\vec{u}\vec{e}_{\sigma})^2}{c^2}
\frac{\vec{k}^2_{\tau}c^2-\omega^2}{(\omega-\vec{k}_{\tau}\vec{u})^2}
,~ \tilde{\varepsilon}_{-\tau}=\varepsilon_{-\tau}=\chi_{-\tau}
\end{eqnarray*}

Further, we shall analyze the generation of a wave with wave
vector  $\vec{k}$, making a small angle with the particle velocity
vector $\vec{u}$. In this case, the wave vector
$\vec{k}_{\tau}=\vec{k}+\vec{\tau}$ is directed at a large angle
relative to $\vec{u}$, and consequently the magnitude of
$(\omega-\vec{k}_{\tau}\vec{u})$ cannot become small. As a result,
the terms containing the expression
$(\omega-\vec{k}_{\tau}\vec{u})$ in their denominators will be
small and can be ignored. We shall also neglect the term
$\omega^2_L/\gamma$ -- this is justified for real beam densities.

It is well known that the equation set (\ref{phys_1.3}) has
nonzero solutions when its determinant equals zero. This defines
the dispersion equation,  which can be written in the form
\begin{eqnarray}
\label{phys_1.4}
(\omega-\vec{k}\vec{u})^2\left[(k^2c^2-\omega^2\varepsilon_0)
(k^2_{\tau}c^2-\omega^2\varepsilon_0)-\omega^4\varepsilon_{\tau}\varepsilon_{-\tau}\right]
=\\
-\frac{\omega^2_L}{\gamma}\frac{(\vec{u}\vec{e}_{\sigma})^2}{c^2}
(k^2c^2-\omega^2)(k^2_{\tau}c^2-\omega^2\varepsilon_0). \nonumber
\end{eqnarray}
Dispersion equation (\ref{phys_1.4}) yields the relationship for
$\omega=\omega(\vec k)$ or $\vec k=\vec k(\omega)$.

%%%%%%%%%%%%%

\section{Generation equations  and threshold conditions in the case of two-wave Bragg diffraction}
\label{laser_sec2}

 In order to determine the structure of the fields
and describe the instability evolution in the systems, one  needs
to know the dispersion equations and their solutions, as well as
the boundary conditions.

Now we shall formulate the boundary problem. Let an electron beam
with mean velocity $\vec{u}$ be incident onto a plane-parallel
spatially periodic plate of thickness $L$. The electron beam is
oriented so that it generates radiation under Bragg diffraction
conditions. Under two-wave diffraction, two fundamentally
different geometries are possible.

\begin{figure}[htpb]
\centering \epsfxsize = 4 cm \centerline{\epsfbox{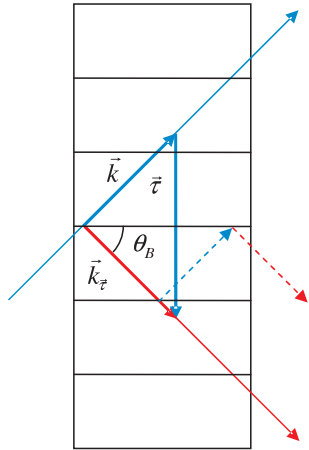}}
\caption{Geometry of two-wave Laue diffraction; $\vec {k},\vec
{k}_{\tau} $ are the wave vectors of the incident and diffracted
waves, respectively,  and $\vec {\tau} $ is the reciprocal lattice
vector of the periodic structure. The projections of both wave
vectors onto the direction of the normal to the surface have the
same sign.} \label{fig.laser_1}
\end{figure}

In the first case (Laue geometry (see Fig.\ref{fig.laser_1}), both
waves are emitted through one and the same boundary of the
periodic structure ($ \gamma _{0} = \frac{{\left( {\vec {k}\vec
{n}} \right)}}{{k}} > 0$ and $\gamma _{1} = \frac{{\left( {\vec
{k}_{\tau } \vec {n}} \right)}}{{k}} > 0$, here $\vec n$ is the
unit normal vector to the entrance surface).

\begin{figure}[h]
\centering
\epsfxsize = 4 cm \centerline{\epsfbox{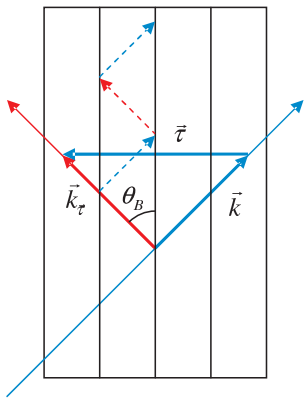}}
\caption{Geometry of  two-wave Bragg diffraction; $\vec {k},\vec
{k}_{\tau}$ are the wave vectors of the incident and diffracted
waves, respectively and  $\vec {\tau}$ is the reciprocal lattice
vector of the periodic structure. The projection of the wave
vectors onto the direction of the normal to the surface are
opposite in sign.} \label{fig.laser_2}
\end{figure}

In the second case (Bragg geometry, see Fig. \ref{fig.laser_2}),
the incident and diffracted waves leave the plate through the
opposite surfaces ($\gamma _{0} > 0$ and  $\gamma _{1} < 0$).

Let us consider the Bragg case in more detail

The general solution for the field in a crystal is written as
\begin{equation}
\label{phys_1.6}
\vec{E}=\sum\limits_{i=1}^4\vec{e}_{\sigma}c_i\exp(i\vec{k_i}\vec{r})[1+s_i\exp(i\vec{\tau}\vec{r})],
\end{equation}
where $\vec{k}_i$ is the i-th solution to dispersion equation
(\ref{phys_1.4}) and $\vec{k}_{i\tau}=\vec{k}_i+\vec{\tau}$,  with
$\vec{\tau}$ being the reciprocal lattice vector corresponding to
the planes of  diffraction reflection. In writing
(\ref{phys_1.6}), we used four roots of dispersion equation
(\ref{phys_1.4}), instead of six. Two roots can be discarded
because at $|\chi_{\tau}|\ll 1$, mirror-reflected waves can be
ignored.

The boundary conditions, necessary for defining the fields in a
crystal,  may be written as \cite{LANL98,bar1}
\begin{eqnarray}
\label{phys_1.7}
& &c_1+c_2+c_3+c_4=1,\nonumber\\
& &f_1c_1+f_2c_2+f_3c_3+f_4c_4=0,\nonumber\\
& &g_1c_1+g_2c_2+g_3c_3+g_4c_4=0,\\
& &s_1c_1e^{iK_{1z}L}+s_2c_2e^{iK_{2z}L}+s_3c_3e^{iK_{3z}L}+s_4c_4e^{iK_{4z}L}=0,\nonumber\\
&
&s_i=\frac{\omega^2\varepsilon_{-\tau}}{k^2_{i_{\tau}}c^2-\omega^2\varepsilon_0},\qquad
f_i=\frac{(\vec{u}\vec{e}_{\sigma})^2}{(\omega-\vec{k}_i\vec{u})},\qquad
g_i=\frac{k^2_ic^2-\omega^2}{(\omega-\vec{k}_i\vec{u})^2}\frac{(\vec{u}\vec{e}_{\sigma})^2}{c^2}.\nonumber
\end{eqnarray}
The first equation in (\ref{phys_1.7}) corresponds to the
continuity of the incident wave at the boundary $z=0$ (the wave
incident on the crystal is assumed to have a unit amplitude). The
last equation in (\ref{phys_1.7}) reflects the fact that behind
the crystal, no waves move in the diffraction direction (they
appear due to diffraction in space before the entrance surface of
the crystal; the general case, which is true for both Bragg and
Laue geometries, is considered in \cite{LANL98}). Here  $i=1 \div
4$ are the solutions to dispersion equation (\ref{phys_1.4}). The
second and the third conditions correspond to the continuity of
the beam density and the beam current density at the crystal
entrance.
Here  we apply the  expressions obtained from the equations of
particle motion and the expression for the particle beam current
\begin{eqnarray}
& & \delta
j_{\sigma}=\frac{ie^2n_0}{m\gamma\omega}\frac{(\vec{u}\vec{e}_{\sigma})^2}{c^2}
\frac{k^2c^2-\omega^2}{(\omega-\vec{k}\vec{u})^2}E_{\sigma}\nonumber\\
& &
j_{\sigma}=e(\vec{u}\vec{e}_{\sigma})n_0-\frac{ie^2n_0}{m\gamma}
\frac{(\vec{u}\vec{e}_{\sigma})^2}{c^2(\omega-\vec{k}\vec{u})}E_{\sigma}.
\label{phys_1.5}
\end{eqnarray}

The linear system (\ref{phys_1.7}), defining the coefficients
$c_i$, has the solution $c_i=\Delta_i/\Delta$, where $\Delta$ is
the determinant of the system (\ref{phys_1.7}) and $\Delta_i$ is
the $i$-th minor, obtained as a result of replacement of the
$i$-th column by
\[\left(
  \begin{array}{c}
    1 \\
    0 \\
    0 \\
    0
  \end{array}
\right).
 \]
 Hence, at $\Delta\rightarrow 0$, the field amplitudes
inside the crystal increase, and as a result, the field in the
crystal is nonzero, even  when the amplitude of the incident wave
vanishes.
 The condition $\Delta=0$  with $\Delta_i\neq 0$ is called
the generation threshold condition \cite{nim95_2}. Substituting
the expressions
\begin{eqnarray}
\label{phys_1.8}
& &\vec{k}_i=\vec{k}_0+\vec{k}\delta_i\vec{n},\qquad k_{0z}=\frac{\omega-\vec{k}_{\perp}\vec{u}_{\perp}}{u_z},\nonumber\\
& &k=\omega/c,\qquad \delta_i\ll 1,
\end{eqnarray}
(where $\vec{n}$ is the normal to the crystal surface and
$\vec{k}_0=(k_{0z},\vec{k}_{\perp})$) into the determinant
$\Delta$, we can represent the generation threshold condition
$\Delta=0$ as
\begin{eqnarray}
\label{phys_1.9} \hspace{-10mm}& &
\frac{(\delta_1-\delta_2)(\delta_1-\delta_3)(\delta_2-\delta_3)}{\delta_1^2\delta_2^2\delta_3^2}s_4e^{ik\delta_4L}
-\frac{(\delta_1-\delta_2)(\delta_1-\delta_4)(\delta_2-\delta_4)}{\delta_1^2\delta_2^2\delta_4^2}s_3e^{ik\delta_3L}\\
\hspace{-10mm}& &
+\frac{(\delta_1-\delta_3)(\delta_1-\delta_4)(\delta_3-\delta_4)}{\delta_1^2\delta_3^2\delta_4^2}s_2e^{ik\delta_2L}
-\frac{(\delta_2-\delta_3)(\delta_2-\delta_4)(\delta_3-\delta_4)}{\delta_2^2\delta_3^2\delta_4^2}s_1e^{ik\delta_1L}=0\,.\nonumber
\end{eqnarray}
In eq. (\ref{phys_1.9}), the terms containing nonresonant  $f_i$
and $g_i$ ($i=1\div 4$) were neglected.

Upon solving  (\ref{phys_1.9}), one may determine the threshold
generation conditions, i.e.,
 the values of the electron current and other parameters of the beam, at which
 radiation begins to exceed the losses \cite{LANL98,bar1}.

 In the case of low-gain regime, for instance, the formula for the generation threshold
under the conditions of two--wave diffraction  was obtained in the
form \cite{bar1}:
\begin{eqnarray}
  &  -\frac{\pi^2 n^2}{4\gamma}\left(\frac{\omega_L}{\omega} \right)^2 k^3 L_*^3
  \left( \chi_0' + |\chi_\tau|/\sqrt{-\beta} - \gamma^{-2} \right)
  \left( \chi_0' + |\chi_\tau|/\sqrt{-\beta} \right)\sin^2\varphi \times f(y)\nonumber  \\
  &  =
  \left( \frac{\gamma_0 c}{\vec{u}\vec{n}} \right)^3
  \frac{16(-\beta)\pi^2 n^2}{(k|\chi_\tau|L_*)^2} + k\chi_0'' L_*
  \left(\frac{1}{\sqrt{-\beta}} - 1\right)^2.
  \label{eq:threshold3D}
\end{eqnarray}
Here  $\omega$ is the radiation frequency;  $\omega_L=\sqrt{4\pi
e^2 n_0/m_e}$ is the Langmuir frequency of the beam; $n_0$ is the
average electron density in the beam;  $e$ and $m_e$ are the
electron charge and mass, respectively; $\gamma$ is the beam
Lorentz factor;  $L_*=Lu/(\vec{u}\,\vec{n})$ is the distance
traveled by the beam in the crystal; $\vec{u}$ is the unperturbed
velocity vector of the beam particles;  $\vec{n}$ is the unit
normal vector to the crystal surface (directed toward the
interior); $L$ is the crystal thickness; $\chi_0$ and $\chi_\tau$
are the Fourier expansion coefficients of the crystal dielectric
susceptibility (their real and imaginary parts are denoted by
prime and double prime, respectively); $\beta=\gamma_1/\gamma_0$
is the diffraction asymmetry factor; $\gamma_0$ and $\gamma_1$ are
the cosines of the angles between the normal vector $\vec{n}$ and
the wave vectors of the transmitted $\vec{k}$ and diffracted
$\vec{k} + \vec{\tau}$ waves, respectively; $\varphi$ is the angle
between the vectors  $\vec{k}_\perp$ and $\vec{\tau}_\perp$ (here
the  subscript $\perp$ denotes the projection of the vector on the
plane perpendicular to $\vec{u}$); $n$ is the integer; $f(y)$ is
the spectral function depending on detuning from the synchronism
conditions:
\begin{equation}
  f(y)=\sin y \frac{(2y+\pi n)\sin y - y(y+\pi n)\cos y}{y^3(y+\pi n)^3},
  \label{eq:spectral}
\end{equation}
where $y = k x_2'L/2$ and  $x_2$ is the root of the dispersion
equation in the absence of the electron beam:
\begin{equation}
  x_{1,2} = a \pm \sqrt{a^2 + b}\,.
  \label{eq:roots1}
\end{equation}
Here
\begin{eqnarray*}
a & = &-\frac{1}{4}(l/\gamma_0 + l_\tau/\gamma_1)\,,\qquad
b  = -\frac{1}{4\gamma_0\gamma_1}(ll_\tau-\chi_\tau\chi_{-\tau})\,,\\
l& = &\theta^2-\chi_0+\gamma^{-2}\,,\qquad\quad l_\tau  =
l+\alpha\,,
\end{eqnarray*}
 where $\theta$ is the angle between the vectors
$\vec{k}$ and $\vec{u}$ and $\alpha=\vec{\tau}(2\vec{k}_0 +
\vec{\tau})/k_0^2$ characterizes deviation from the exact Bragg
condition.

Let us note that  (\ref{eq:threshold3D}) is obtained in the
vicinity of the condition imposed on the phase difference between
the waves in the crystal
\begin{equation}
\label{new}
 \texttt{Re}\,(k_{1z}-k_{2z})L= 2\pi n,
 \end{equation}
 which in the considered case has the form
\begin{equation}
kL\sqrt{a^2+b}=\pi n.
  \label{eq:phasecond}
\end{equation}
 and it is assumed that the following inequalities hold:
\[
k|\chi_\tau|L \gg 1\,, \qquad |\chi_\tau| \ll 1\,, \qquad |\chi_0|
\ll 1\,.
\]
 The condition  (\ref{eq:threshold3D}) has a clear physical meaning:
 the left-hand side of  (\ref{eq:threshold3D}) contains the term
 describing generation of radiation by the electron beam, and the
 right-hand side includes the terms describing losses at the
 boundaries (the first term) and absorption losses (the second
 term) in the medium. It is obvious that if the absorption is
 small, then at  fixed values of $y$ (e.g. if  $y=0$, then f(y)=1), eq. (\ref{eq:threshold3D})
yields the following dependence of the threshold generation
current on the target length (the left-hand side of
(\ref{eq:threshold3D}) is proportional to $\omega_L^2\sim n_0$,
i.e., $\sim j$):
\begin{equation}
  j_{th} \sim  \frac{1}{(kL_*)^5}.
  \label{eq:thresholdL}
\end{equation}
It also follows from (\ref{eq:threshold3D}) that the value of the
threshold current in the case of a non-one-dimensional DFB depends
appreciably on the parameter $\beta$ and the  effective photon
path length  $L_*$  in the resonator (recall that $L_*=L\,u/(\vec
u\vec n)$).

%%%%%%%%%%%%%%%%%%%%%%%%%%%%%%%%%%%%%%%%%%%%%%%%

\section{VFEL based on an electromagnetic undulator}
\label{vesti_sec:1}

Similar considerations enable finding  generation thresholds for
the beams moving in an electromagnetic undulator
\cite{bdz_phl,bdz_vesti,bdz_nim}.

Let an electron beam move with average velocity $\vec{u}_0$
through a two- or three-dimensional periodic medium
(electromagnetic (photonic) crystal) placed in an external field.
Vector potential of the external field can be written as
\begin{enumerate}
\item
\[
\vec{A}=A_w\vec{x}\sin k_wz\qquad \mbox{(a magnetostatic
wiggler)}\,,
\]
where $A_w=B_w/k_w$, $k_w=2\pi/\lambda_w$, and
$\vec{x}\cdot\vec{u}_0=0$, with $B_w$ being the magnetic field
strength and  $\lambda_w$ being the wiggler wavelength ($\vec{x}$,
$\vec{y}$, and $\vec{z}$ are the unit vectors of the Cartesian
coordinate system)($c=1$); \item
\[
A=A_{em}\vec{x}\sin (\vec{k}_{em}\vec{r}-\omega_{em}t)\,,
\]
where $A_{em}=E_{em}/\omega_{em}$. Here $E_{em}$ is the amplitude
of the electric field of the electromagnetic wave and
$\omega_{em}$,  $\vec{k}_{em}$ are its frequency and wave vector,
respectively. The external field excites oscillations of
electrons,  having the transverse velocity
$\vec{v}_{\perp}=e\vec{A}/m\gamma$, where $\gamma= (1-v^2)^{-1/2}$
is the Lorentz factor of the electron. A moving oscillator emits
electromagnetic radiation, whose frequency $\omega$ and wave
vector $\vec{k}$ satisfy the condition of synchronism:
\begin{eqnarray}
\label{1} \noindent\omega-\vec{k}\vec{u}_{\parallel}=\Omega,
\Omega =\left\{\begin{array}{l}\vec{k}_w\vec{u}_{\parallel}
\quad(\mbox{for a magnetic
wiggler})\,,\\
\\
\omega_{em}-\vec{k}_{em}\vec{u}\quad(\mbox{for an electromagnetic
wiggler})\,,
\end{array}\right.
\end{eqnarray}
where the longitudinal velocity of the electron
$u_{\parallel}=u_0\left(1-\frac{a^2_w}{\gamma^2}\right)$ (the
corresponding longitudinal Lorentz factor
$1/\gamma^2_{\parallel}=(1+a^2_w)/\gamma^2$, $a_w=eA_w/m$ is the
undulator parameter; $k=n(\omega)\omega$,
$k_{em}=n(\omega_{em})\omega_{em}$, and $|n(\omega)-1|\ll 1$). We
shall further consider the case of forward scattering.
\end{enumerate}

 Let us
write Maxwell's equation in the Fourier space in the two-wave
approximation of the dynamical diffraction theory (when the Bragg
condition is satisfied only for one reciprocal lattice vector).
For the field of  $\sigma$-polarization, we have:
\begin{equation}
\label{2}
[k^2-\omega^2(1+\chi_0)]E-\omega^2\chi_{-\tau}E_{\tau}=4\pi
i\omega j\,,
\end{equation}
\begin{equation}
\label{3}
-\omega^2\chi_{\tau}E+[k_{\tau}^2-\omega^2(1+\chi_0)]E_{\tau}=4\pi
i\omega j_{\tau}\,,
\end{equation}
where $E=\vec{E}(\vec{k},\, \omega)\cdot\vec{e}_{\sigma}$ and
$E_{\tau}=\vec{E}(\vec{k}_{\tau}\,,\omega)\cdot\vec{e}_{\sigma}$
are the Fourier components of the electromagnetic field of
$\sigma$-polarization, $\vec{k}_{\tau}=\vec{k}+\vec{\tau}$,
$\vec{e}_{\sigma}=\vec{k}\times\vec{k}_{\tau}/|\vec{k}\times\vec{k}_{\tau}|$
is the $\sigma$-polarization unit vector, $j=\vec{j}(\vec{k}\,,
\omega)\cdot\vec{e}_{\sigma}$ and
$j_{\tau}=\vec{j}(\vec{k}_{\tau}\,, \omega)\cdot\vec{e}_{\sigma}$
are the projections of the Fourier components of the current
density vector onto the $\sigma$-polarization unit vector, and
$\chi_0$ and  $\chi_{\tau}$ are the Fourier components of the
periodic tensor of crystal polarizability.

Let us use the linear approximation for solving the equations
\begin{equation}
\label{4} \frac{\partial f}{\partial t}+\vec{v}\frac{\partial
f}{\partial \vec{r}}+\frac{\partial \vec{p}}{\partial
t}\,\frac{\partial f}{\partial \vec{p}}=0\,,\qquad \frac{\partial
\vec{p}}{\partial t}=e(\vec{E}+\vec{v}\times\vec{B})\,.
\end{equation}
In this case, we can write
\begin{equation}
\label{5} f(\vec{r}\,, \vec{p}\,, t)=f_0(\vec{p})+\delta
f(\vec{r}\,, \vec{p}\,, t),\quad \delta f \sim E,\quad \delta f
\ll f_0\,.
\end{equation}

  From (\ref{4}) and  (\ref{5})  follows
  \begin{equation}
  \label{6}
  \left(\frac{\partial}{\partial
  t}+\vec{v}\frac{\partial}{\partial\vec{r}}\right)\delta
  f+\frac{\partial\vec{p}}{\partial t}\, \frac{\partial
  f_0}{\partial \vec{p}}=0\,.
  \end{equation}
In the equation of motion (\ref{6}) and the expression
$\vec{j}=e\int\vec{v} f(\vec{r}\,,\vec{p}\,,t) d^3 p $ for the
current density, we shall go over to the Fourier components. Let
us express the perturbation of the distribution function in
(\ref{6}) in terms of $E(\vec{k}\,, \omega)$ and then substitute
it into (\ref{2}) and (\ref{3}). If the beam's velocity
distribution $v_{th}$ satisfies the relationships
$\vec{k}\cdot\vec{v}_{th}\ll \Omega$ and
$\vec{k}\cdot\vec{v}_{th}\ll \vec{\tau}\cdot\vec{u}$, then one can
obtain the dispersion equation in the form:
\begin{eqnarray}
\label{7} D_0(\vec{k}, \omega)=-4\pi
e^2\omega(k^2_{\tau}-\omega^2(1+\chi_0))\frac{a^2_w}{\gamma^2}I\,,\nonumber\\
D_0(\vec{k},
\omega)=(k^2-\omega^2(1+\chi_0))(k^2_{\tau}-\omega^2(1+\chi_0))-\omega^4\chi_{\tau}\chi_{-\tau}\,,
\end{eqnarray}
\begin{equation}
\label{8}
I=\int\frac{df_0}{dp}\,\frac{dp}{\vec{k}\vec{u}_{\parallel}-\omega+\Omega}\,,\qquad
p=\vec{p}\cdot\vec{s}\,,\qquad
\vec{s}=\frac{\vec{u}_{\parallel}}{u_{\parallel}}\,.
\end{equation}
Let a crystal have a form of a plane-parallel plate of thickness
$L$, then $L_*$ denotes the particle path length in the crystal
(see Sec. \ref{laser_sec2}). Here we shall consider only the case
of weak amplification, when $k^{\prime\prime}L_*\ll 1$, with
$k^{\prime\prime}$ being the imaginary part of the solution to
dispersion equation (\ref{7}). There are two characteristic
limiting cases for which (\ref{7}) can be reduced to the algebraic
equation.
\begin{enumerate}
\item If the inequality
\begin{equation}
\label{9} \vec{k}\vec{v}_{th}\ll 1/L_*
\end{equation}
is fulfilled, then we can neglect the beam dispersion and set the
initial distribution function to be $f_0(p)=n_0\delta(p-p_0)$,
where $n_0$ is the beam's spatial density. As a result, we can
write the dispersion equation for a cold-beam regime.
\begin{equation}
\label{10} D_0(\vec{k}\,,
\omega)(\omega-\vec{k}\vec{u}_{\parallel}-\Omega)^2=-2\omega^4(k^2_{\tau}-\omega^2(1+\chi_0))F\,.
\end{equation}
Here $F=\frac{\omega^2_p
a_w^2}{2\gamma^3\gamma^3_{\parallel}\omega^2}$ and
$\omega^2_p=\frac{4\pi e^2 n_0}{m}$ is the plasma frequency of the
beam.
 \item
If inequality (\ref{11}), inverse to (\ref{9}),
 is fulfilled:
 \begin{equation}
 \label{11}
 \vec{k}\vec{v}_{th}\gg 1/L_*\,,
 \end{equation}
  then we can speak of a "hot" beam regime. To evaluate the integral in (\ref{8}), we use the Gaussian
  distribution function
  $f_0(p)=\frac{n_0}{\sqrt{\pi}p_{th}}e^{-(p-p_0)^2/p^2_{th}}$ and
  obtain the dispersion equation in the "hot" beam regime:
  \begin{eqnarray}
  \label{12}
  D_0(\vec{k}\,,
  \omega)&=&-2\omega^2(k^2_{\tau}-\omega^2(1+\chi_0))\frac{F}{v^2_{th}}\,\frac{2}{\sqrt{\pi}}\left(1+i\sqrt{\pi}\frac{u_f-u_{\parallel
  0}}{v_{th}}\right)\,,\\
  u_f &=&\left\{\begin{array}{l}
  \frac{\omega}{k_{\parallel}+k_w}\, \mbox{(for a magnetic
  wiggler)}\,,\\
\frac{\omega-\omega_{em}}{k_{\parallel}+k_{em}}\quad \mbox{(for an
electromagnetic   wiggler)}\,.\nonumber
\end{array}\right.
\end{eqnarray}
\end{enumerate}

\subsection{Boundary Conditions}

Let a wave vector of the incident wave be  $\vec{k}_0=\omega
\vec{s}$. From the continuity of the wave field at the boundary
follows that inside the crystal, the wave vector
$\vec{k}=\vec{k}_0+\omega \delta\vec{n}$ ($\delta\ll 1$ because
$|\chi_0|$, $|\chi_{\tau}|\ll 1$). Let us introduce the following
notations: $\gamma_0=\vec{k}_0\cdot\vec{n}/\omega$,
$\gamma_1=(\vec{k}_0+\vec{\tau})\cdot\vec{n}/\omega$,
$\beta_1=\gamma_0/\gamma_1$,
$\alpha=((\vec{k}_0+\vec{\tau})^2-\omega^2)/\omega^2$,
$\chi_1=\chi_0-\alpha$, $\varepsilon=\gamma_0\delta$, and
$\xi=-\frac{1}{2\gamma^2_{\parallel}}-\frac{F}{\omega}$ and recast
dispersion equations (\ref{10}), (\ref{12}) in a more convenient
form:
\begin{equation}
\label{13}
(\varepsilon-\varepsilon_1^{(0)})(\varepsilon-\varepsilon_2^{(0)})(\varepsilon-\varepsilon_k)=
-\left(\varepsilon+|\beta_1|\frac{\chi_1}{2}\right)F\,,
\end{equation}
\begin{equation}
\label{14}
(\varepsilon-\varepsilon_1^{(0)})(\varepsilon-\varepsilon_2^{(0)})=-\left(\varepsilon+|\beta_1|\frac{\chi_1}{2}\right)\frac{F}{v^2_{th}}\,
\frac{2}{\sqrt{\pi}}\,
\left(1+i\sqrt{\pi}\frac{(u_f-u_0)}{v_{th}}\right)\,,
\end{equation}
\begin{eqnarray}
\label{15}
\varepsilon_1^{(0)}=\frac{1}{4}\left[\chi_0-|\beta_1|\chi_1+\sqrt{(\chi_0+|\beta_1|\chi_1)^2-4|\beta_1|\chi_{\tau}\chi_{-\tau}}\right]\,,\nonumber\\
\varepsilon_2^{(0)}=\frac{1}{4}\left[\chi_0-|\beta_1|\chi_1-\sqrt{(\chi_0+|\beta_1|\chi_1)^2-4|\beta_1|\chi_{\tau}\chi_{-\tau}}\right]\,.
\end{eqnarray}
In writing (\ref{13}) and  (\ref{14}), we have taken into account
that in the case of Bragg geometry considered here (the diffracted
wave leaves the crystal through the entrance surface),
$\beta_1<0$. The general solution describing scattering of the
external wave with the field strength
$\vec{E}=E_0\vec{x}e^{i(\vec{k}_0\vec{r}-\omega t)}$ by the system
"crystal-beam" is as follows:
\begin{equation}
\vec{E}=\vec{x}e^{i(\vec{k}_0\vec{r}-\omega
t)}\sum_{j=1}^{N}(E_j+e^{i\vec{\tau}\vec{r}}E_{\tau
j})e^{i\varepsilon_j\omega z/\gamma_0}\,,
\end{equation}
and, according to (\ref{3}), $\vec{E}_{\tau
j}=\vec{E}_j\frac{\chi_{\tau}}{2\varepsilon_j+|\beta_1|\chi_1}$;
$N=4$ and $N=2$ for a "cold" and a "hot" beam, respectively.  To
determine the amplitudes of the fields inside the crystal, the
boundary conditions must be used. In Bragg geometry, they are as
follows:
\begin{enumerate}
\item the continuity of field $E$ at the entrance boundary
\begin{equation}
\label{17} \sum_{j=1}^{N} E_j=E_0\,,
\end{equation}
\item the continuity of field $E_{\tau}$ at the exit boundary
\begin{equation}
\label{18} \sum_{j=1}^{N}\frac{E_j e^{i\varepsilon_j\omega
L}}{\varepsilon_j+|\beta_1|\frac{\chi_1}{2}}=0\,,
\end{equation}
\item the continuity of the beam's density at the entrance
boundary
\begin{equation}
\label{19} \int\delta f dp|_{z=0}=0\,,
\end{equation}
\item the continuity of the beam's current density  at the
entrance boundary
\begin{equation}
\label{20} \int u\delta f dp|_{z=0}=0\,.
\end{equation}
\end{enumerate}
Using the expression for the Fourier components $df$, which
follows from (\ref{6}), let us express (\ref{19}) and (\ref{20})
in terms of the amplitude $E_j$ in the crystal. It can be shown
that for the case of a "cold" beam, expressions (\ref{19}) and
(\ref{20}) yield the equations:
\begin{equation}
\label{21} \sum_{j=1}^{4}\frac{E_j}{\xi -
\varepsilon_j}=0\,,\qquad \sum_{j=1}^{4}\frac{E_j}{(\xi -
\varepsilon_j)^2}=0\,.
\end{equation}
These equations, together with equations (\ref{17}) and
(\ref{18}), give four independent conditions that are necessary to
define the four amplitudes $E_j$ of the field. For a "hot" beam,
both of the conditions (\ref{19}) and  (\ref{20}) can be reduced
to (\ref{17}), i.e., only the conditions (\ref{17}) and (\ref{18})
are independent and sufficient for  defining the two amplitudes of
the field. Let $g_{ij}$ denote the elements of the fundamental
matrix of the system (\ref{17}), (\ref{18}), and (\ref{21}). The
boundary conditions enable one to determine the field  amplitudes,
which can be written in the form $E_j=E_0\bar{g}_{j1}/G$, where
$\bar{g}_{j1}$ is the cofactor of the element $g_{j1}$, and
$G=\det \|g_{i1}\|$. If $G$ vanishes, then the amplitudes $E_j$ in
the crystal can be nonzero, even when the amplitude $E_0$ of the
external incident field equals zero. This correlates with the
phenomenon of self-excitation (generation), arising due to the
existence of the distributed feedback. On this account, we shall
call the equation $G=0$ the generation condition. Calculating the
determinant of the matrix $\|g_ij\|$, one can reduce the
generation condition to the form:
\begin{equation}
\label{22} \sum_{j=1}^{4}\frac{e^{i(\varepsilon_j-\xi)\omega
L}(\varepsilon_j-\xi)^2}{\left(\varepsilon_j+|\beta_1|\frac{\chi_1}{2}\right)\prod_{i\neq
j}(\varepsilon_i-\varepsilon_j)}=0
\end{equation}
for a "cold" beam and
\begin{equation}
\label{23} \frac{e^{i\varepsilon_1\omega L_*}}{\varepsilon_1+
|\beta_1|\frac{\chi_1}{2}}=\frac{e^{i\varepsilon_2 \omega
L_*}}{\varepsilon_1+ |\beta_1|\frac{\chi_1}{2}}
\end{equation}
for a "hot" beam.

\subsection{Generation Thresholds}

Recall that the Fourier components of crystal polarizability have
the form:
\[
\chi_0=-|\chi_0^{\prime}| + i|\chi_0^{\prime\prime}|\,,\quad
\chi_{\pm\tau}=-|\chi_{\pm\tau}^{\prime}| +
i|\chi_{\pm\tau}^{\prime\prime}|\,,\quad
|\chi_0^{\prime\prime}|\,,
|\chi_{\pm\tau}^{\prime\prime}|\ll|\chi_0^{\prime}|\,,|\chi_{\tau}^{\prime}|.
\]

Now, let us analyze the case when
\begin{equation}
\label{25}
\sqrt{|\chi_{0}^{\prime\prime}|\,|\chi_{\tau}^{\prime}|}\ll
\varepsilon_1^{(0)}-\varepsilon_2^{(0)}\ll
|\chi_{\tau}^{\prime}|\,.
\end{equation}
As is seen from (\ref{15}), this occurs at small deviations from
$\alpha \simeq\frac{1}{|\beta_1|}(\pm 2
\sqrt{|\beta_1|}|\chi_{\tau}^{\prime}|-|\chi_{0}^{\prime}|(1+|\beta_1|))$,
i.e., in the vicinity of the absorption edges.  The solutions to
(\ref{13}) at $\texttt{Re}\,\varepsilon_1^{(0)}=\xi$ or
$\texttt{Re}\,\varepsilon_2^{(0)}=\xi$ are of particular interest
and correspond  to the synchronism between the beam and one of the
 DFB resonator modes. In Table I, the possible cases are
labelled with numbers and letters for convenience.

\begin{table}[ht]\center
\caption{}
\begin{tabular} {c|c|c}
\small{Synchronism condition} &
\scriptsize{$\alpha=\frac{1}{|\beta_1|}\left(2\sqrt{|\beta_1|}|\chi^{\prime}_{\tau}|-
|\chi^{\prime}_{0}|(1+|\beta_1|)\right)$} & \scriptsize{$\alpha
=\frac{1}{|\beta_1|}\left(-2\sqrt{|\beta_1|}|\chi^{\prime}_{\tau}|-
|\chi^{\prime}_{0}|(1+|\beta_1|)\right)$}\\
$\texttt{Re}\,\varepsilon_1^{(0)}=\xi$ & 1a) & 2a)\\
$\texttt{Re}\,\varepsilon_2^{(0)}=\xi$ & 1b) & 2b)
\end{tabular}
\end{table}

In cases 1a) and 2b),  equation (\ref{22}) does not have a
solution.

The condition (\ref{22}) can be satisfied when the
following equalities hold:\\
in case 1b)
\begin{equation}
\label{26} \frac{F(\omega L_*)^3}{8 \pi n}=
\omega\left(|\chi_{0}^{\prime\prime}|\frac{(1+|\beta_1|)}{2}-|\chi_{\tau}^{\prime\prime}|\sqrt{|\beta_1|}\right)L_*+\frac{8
\pi^2 n^2}{\omega^2\chi_{\tau}L_*^2}\,;
\end{equation}
in case 2a)
\begin{equation}
\label{27} \frac{F(\omega L_*)^3}{8 \pi n}=
\omega\left(|\chi_{0}^{\prime\prime}|\frac{(1+|\beta_1|)}{2}+|\chi_{\tau}^{\prime\prime}|\sqrt{|\beta_1|}\right)L_*+\frac{8
\pi^2 n^2}{\omega^2\chi_{\tau}L_*^2}\,.
\end{equation}
Let us recall that $F=\frac{\omega^2_p
a_w^2}{2\gamma^3\gamma^3_{\parallel}\omega^2}$.  The solutions to
(\ref{23}) in the case of a "hot" beam have the form
\begin{enumerate}
\item
\begin{eqnarray}
\label{28} \omega\frac{F}{v^3_{th}}(u_{\parallel\, 0}-u_f)L_* & =
&
\omega\left(|\chi_{0}^{\prime\prime}|\frac{(1+|\beta_1|)}{2}-|\chi_{\tau}^{\prime\prime}|\sqrt{|\beta_1|}\right)L_*+\frac{8
\pi^2 n^2}{\omega^2\chi_{\tau}L_*^2}\,,\nonumber\\
\alpha & = &
\frac{1}{|\beta_1|}\left(2\sqrt{|\beta_1|}|\chi^{\prime}_{\tau}|-
|\chi^{\prime}_{0}|(1+|\beta_1|)\right)\,, \nonumber\\
 n & = & 1,\, 2\,...
\end{eqnarray}
\item
\begin{eqnarray}
\label{29} \alpha
&=&\frac{1}{|\beta_1|}\left(-2\sqrt{|\beta_1|}|\chi^{\prime}_{\tau}|-
|\chi^{\prime}_{0}|(1+|\beta_1|)\right)\,, \nonumber\\
\frac{\omega F}{v^3_{th}}(u_{\parallel\, 0}-u_f)L_* &=&
\omega\left(|\chi_{0}^{\prime\prime}|\frac{(1+|\beta_1|)}{2}-|\chi_{\tau}^{\prime\prime}|\sqrt{|\beta_1|}\right)L_*+\frac{8
\pi^2 n^2}{\omega^2\chi_{\tau}L_*^2}\,,\nonumber\\
n &=& 1,\, 2\,...
\end{eqnarray}
\end{enumerate}
Since $F\sim \omega_p^2\sim n_0\sim j$, the value of the threshold
current can be found using  equations (\ref{26})--(\ref{29}).

 Again, we have
$j_{th}\sim \frac{1}{L_*^5}$ for a "cold" beam. Equations
(\ref{26})--(\ref{29}) are the amplitude conditions of generation.
To satisfy the generation conditions, one more condition must be
fulfilled - the phase condition, common for all the cases:
\begin{equation}
\label{30} \texttt{Re}\,
\varepsilon_1^{(0)}-\varepsilon_2^{(0)}=2\pi n/\omega L_*\,.
\end{equation}

The condition (\ref{30}) leads to the fact that the longitudinal
structure of the eigenmodes $|E|^2$ and  $|E_{\tau}|^2$ appears to
be close in form to a standing wave. Thus, this condition is
similar to a well-known condition  of standing wave formation in a
mirror resonator.

Formulas (\ref{26})--(\ref{29}) are similar in form. On the
left-hand sides, they contain the gain of the free electron laser
in the appropriate mode. Their right-hand sides describe the
losses related to energy absorption in crystals and its leakage
through the boundaries of a DFD resonator.

%%%%%%%%%%%%%%%%%%%%%%%%%%%%%%%%%%%%%%%%%%%%%%%%
It should be noted that in view of the above, the value of the
threshold current in the case of a non-one-dimensional DFB depends
appreciably on the parameter $\beta_1$ and the  effective photon
path length  $L_*$  in the resonator (recall that $L_*=L\,u/(\vec
u\vec n)$).

Let us consider the equation defining the threshold conditions for
backward Bragg diffraction ($|\beta|, |\beta_1|=1$ and $\vec
u\parallel\vec n$). In this case, we go over to a one-dimensional
DFB. The implicit form of the equation for the threshold current
was obtained in \cite{E2} (see eq. (10) in \cite{E2}). From this
equation, one can also derive the expressions for the threshold
current. To do this, let us take eqs. (9) and (10), derived in
\cite{E2}, and
 substitute  eq. (9) for the quality factor $Q$  into eq.
(10). Then we can write
\begin{equation}
  (CL')^3 \varphi'(\Phi) = \frac{2\pi^2 n^2}{(\sigma L')^2},
  \label{eq:thGinzburg}
\end{equation}
where $n$ is the integer; $L'$ is the dimensionless length of the
structure, e.g., the waveguide with shallow corrugations
 ($L'=\frac{\omega}{c}L$); $\sigma$ is the so-called wave coupling
 coefficient due to corrugations, whose physical meaning is similar to that of the quantity  $\chi_\tau$. Here $C=(eI\kappa^2\mu/m\gamma\omega^2
N_s)^{1/3}$ is the generalized Pierce parameter, where   $I$ is
the current of the electron beam, $\kappa$ is the coupling
coefficient between the direct wave and and the electron beam,
$\mu$ is the electron bunching parameter, and  $N_s$ is the norm
of the wave in a smooth wavequide. In (\ref{eq:thGinzburg}),
$\Phi=-\Gamma_e L'$, where $\Gamma_e=-(\Delta +\delta)$. Here
$\Delta=(\omega-\omega_B)/\omega_B \approx (h-h_B)/h_B$ is
detuning from the exact Bragg condition; $h_B=\pi/d$, $\omega_B=
c\sqrt{g^2+ h_B^2}$, $d$ is the corrugation period, $h$ and $g$
are the longitudinal and  transversal wavenumbers, respectively,
and $\delta$ is the resonance detuning between the synchronous
wave and the electrons.

The function $\varphi'(\Phi)$ has the form:
\begin{equation}
  \varphi'(\Phi) \equiv \frac{d}{d\Phi}\left\{2\pi^2 n^2 \frac{1 - (-1)^n\cos\Phi}{(\Phi^2-\pi^2
  n^2)^2}\right\}\,.
    \label{eq:specGinzburg}
\end{equation}
Now let us write the explicit form of the derivative of
$\varphi'(\Phi)$
\begin{equation}
\label{eq:specGinzburg1}
  \varphi'(\Phi)  =  2\pi^2 n^2 \frac{(-1)^n\sin\Phi (\Phi^2 -\pi^2
n^2) - 4\Phi (1 - (-1)^n\cos\Phi)}{(\Phi^2 - \pi^2 n^2)^3}\,.
\end{equation}
 At
first glance, the spectral functions in   (\ref{eq:spectral}) and
(\ref{eq:specGinzburg}) differ. It can be shown, however, that
virtually  they are one and the same function, but of different
arguments. Indeed, in the case of a symmetric backward Bragg
diffraction, we have  $\beta\approx -1$, and the quantity
 $a$ in (\ref{eq:roots1}) is  $a =\alpha/4 \approx
\Delta$. Then assuming that $\delta=0$ and considering the
condition  (\ref{eq:phasecond}), we get $2y \approx \Phi + \pi n$.
Substituting this equality for the arguments into
(\ref{eq:specGinzburg}), we obtain $\varphi'(\Phi) =
\varphi'(2y+\pi n)=-\frac{\pi^2 n^2}{4}f(y)$.  Thus the condition
 (\ref{eq:thGinzburg}) takes the form:
\begin{equation}
  -\frac{\pi^2 n^2}{4}(CL')^3 f(y) = \frac{2\pi^2 n^2}{(\sigma L')^2},
  \label{eq:thGinzburg2}
\end{equation}
which is equivalent to (\ref{eq:threshold3D}) (note, however, that
in (\ref{eq:thGinzburg2}), the  radiation absorption is ignored).
Recalling that $C^3\sim I$, we immediately obtain that  the
condition (\ref{eq:thGinzburg2}) yields the same dependence of the
threshold current density on $L$ as in eq. (\ref{eq:threshold3D}).

We may note in conclusion that in contrast to a one-dimensional
case,  where  Laue geometry  does not occur, in a
non-one-dimensional case it occurs along with the generation in
Bragg geometry. A more detailed analysis of the radiation process
in Laue geometry can be found in \cite{LANL98} and the references
there.

\section{Generation equations and threshold conditions in the geometry of three-wave Bragg diffraction}
\label{laser_sec3}

Above, we have discussed the theory of the volume distributed
feedback formation in the geometry of a non-one-dimensional
two-wave Bragg diffraction.   Let us proceed now to the
consideration of multi-wave DFB in a two(three)-dimensional
periodic resonator. Multi-wave DFB arises when $N$ number of waves
simultaneously satisfy the conditions of Bragg diffraction.

Use of multi-wave Bragg diffraction in a VFEL for the formation of
a volume distributed feedback enables one, on the one hand, to
appreciably reduce the length of the generation area  (at a given
operating current) or the operating current (at a given length of
the generation area), and on the other hand, to apply electron
beams with a large transverse cross section (or several electron
beams) for generating radiation, which improves the electrical
endurance of the generator.

Application of multi-wave diffraction for generating in a
microwave range has one more remarkable feature -- the possibility
of selecting modes in oversized waveguides and resonators.

Production of  high-power microwave pulses requires high electric
strength of the generator and radiation resistance of the output
window.  To reduce the load on these elements, the transversal
(with respect to the direction of the electron beam velocity)
dimension of the resonator should be large (much larger than the
wavelength).  As a rule, this leads to a multi-mode  generation
regime and low efficiency.
When $N$ number of waves are diffracted, the modes can be
effectively selected  due to the requirement to satisfy the Bragg
condition $N-1$ .

\begin{figure}[htbp]
\centering
\epsfxsize = 5 cm \centerline{\epsfbox{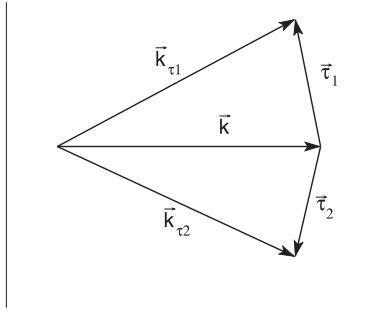}}
\caption{Three-wave diffraction in Laue-Laue geometry. Projections
of wave vectors $\vec {k},\vec {k}_{\tau 1}$, and $\vec {k}_{\tau
2} $  onto the surface normal $\vec n$  have the same sign; $\vec
{\tau} _{1}$ and $\vec {\tau} _{2} $ are the reciprocal lattice
vectors of the periodic structure.} \label{fig.laser_5}
\end{figure}

  To illustrate the potential of  multi-wave distributed feedback,
  let us  consider  three-wave diffraction in more depth (for details, see Section 13 in
  \cite{LANL98}).
  In this case, the  distributed feedback  can be realized in three different geometries:

(1) Laue-Laue diffraction when the three  waves exit through the
same surface ($\gamma _{0} ,\gamma _{1} ,\gamma _{2} > 0$,
 here $\gamma_i $ is the cosine of the angle between the wave
vector $\vec {k}_i$ and the normal to the crystal surface (see
e.g. \cite{132}); Fig. \ref{fig.laser_5});

(2) diffraction in  Bragg-Bragg geometry when $\gamma _{0} > 0$,
while  $\gamma_{1} ,\gamma _{2} < 0$;

\begin{figure}[htbp]
\centering
\epsfxsize = 5 cm \centerline{\epsfbox{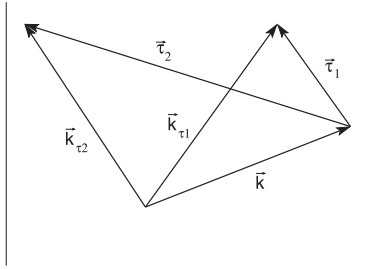}}
\caption{Three-wave diffraction in  Laue-Bragg geometry.
Projections of wave vectors $\vec {k}$ and $\vec {k}_{\tau 1} $
onto the surface normal are opposite in sign to the projection of
vector $\vec{k}_{\tau 2}$; $\vec {\tau} _{1}$ and $\vec {\tau}
_{2} $ are the reciprocal lattice vectors of the periodic
structure.} \label{fig.laser_6}
\end{figure}

(3) Bragg-Laue diffraction when  $\gamma _{0} ,\gamma _{1} > 0$,
and  $\gamma _{2} < 0$; see Figure \ref{fig.laser_6}.

Similarly to the two-wave case, the problem of the beam
interaction with the resonator (photonic crystal) can be reduced
to the problem of three-wave diffraction of an electromagnetic
wave incident onto the active medium. The active medium here is
the system "spatially--periodic structure + electron beam".

%%%%%%%%%%%%%%%%%
The dependence of the threshold conditions on the length of the
interaction area at the point of threefold degeneration changes
appreciably
\begin{equation}
\label{laser_eq32} G = A\chi _{0}^{\prime\prime} +
\frac{{B}}{{kL}}\left( {\frac{{2\pi} }{{klL}}} \right)^{4}\,.
\end{equation}
Here in the regime of a "cold" electron beam, $G\sim jL^{2}$, with
$j$ being the beam current density.
 Realization of this regime
requires the fulfillment of the phase condition $\left( {k_{1z} -
k_{2z}} \right)L = 2\pi n$, $\left( {k_{2z} - k_{3z}} \right)L =
2\pi m$, $n \ne m$.  The coefficients $A$ and $B$ in
(\ref{laser_eq32}) depend on the diffraction geometry, the value
of $\chi_{\tau}$,  and the indices $m$ and  $n$.
According to (\ref{laser_eq32}), for a "cold" electron beam, when
absorption is not important, $ j_{\mathrm{thr}}
\sim\frac{{1}}{{\left( {kL} \right)^{3}}}\left( {\frac{{2\pi}
}{{klL}}} \right)^{4} $.
Under  dynamical diffraction, when the inequality $ \frac{{4\pi}
}{{klL}} \ll 1 $ holds, this dependence leads to an appreciable
reduction in the threshold current. The analysis shows that when a
multi-wave ($N$-wave) DFB is excited, the threshold current
depends on $L$ as
\begin{equation}
\label{laser_eq33} j_{thr} \sim\frac{1}{\left( kL \right)^3}\left(
\frac{2\pi }{klL} \right)^{2s}\,,
\end{equation}
that is,
\[
j_{th}\sim \frac{1}{L^{3+2 s}}\,,
\]
 where $s=N-1$ is the number of extra waves appearing
through diffraction in the crystal.

So,  the transition to multi-wave diffraction enables one to
significantly reduce the operating current and the longitudinal
dimensions of the generating system.

As follows from the above results, the volume distributed feedback
(non-one-dimensional feedback) has a number of advantages that
make its application beneficial for generating stimulated
radiation in a wide spectral range (with wavelengths from
microwave and optical to angstr\"{o}m).
Moreover, in a short-wave spectral range, where the requirements
for the current density and the quality of the beam are very
strict, it becomes possible to noticeably reduce the threshold
current for a given beam propagation area. In this case, the VFEL
is a unique system providing lasing at relatively small
interaction lengths. When producing high-power radiation pulses in
oversized generators in a microwave range,  VFELs are beneficial
for the reduction of the threshold current, generator size,  and
for mode selection.

%%%%%%%%%%%%%%%%%%%%%%%%%%%%%%%%%%%%%%%%%%%%%%%%%%%%%

\section{Use of a dynamical undulator mechanism to produce short wavelength
radiation in VFELs} \label{mop_sec1}

Numerous applications can benefit from the development of powerful
electromagnetic generators with frequency tuning in millimeter,
sub-millimeter and terahertz wavelength range, using
low-relativistic electron beams.
Low-relativistic electron beams in the undulator system can be
used for radiating in a short-wavelength range, but in this case
the undulators period must be small. For example, to obtain
radiation with the wavelength of $ 0.3 - 1~$ mm at the beam energy
$E=800$ KeV$ - 1 $ MeV, the undulator period must be $\sim
0.3-1~$cm. Development of such undulators  is a very complicated
task. The use of a two-stage FEL with a dynamical wiggler
generated by the electron beam \cite{laser_1} is a possible
solution to this problem.

The above-described potentialities to significantly reduce the
threshold currents and resonator dimensions by using a
non-one-dimensional distributed feedback, which arises due to
Bragg diffraction in a two-- or three--dimensional
spatially-periodic resonator (electromagnetic (photonic) crystal)
and serves as a foundation for designing Volume Free Electron
Lasers (VFELs), enable the development of two-stage VFELs with a
dynamical wiggler generated by electron beams.
A dynamical wiggler can be created with the help of any radiation
mechanism: Cherenkov, Smith-Purcell, quasi-Cherenkov \cite{bar1},
or undulator. VFEL principles  offer the advantage of a two-stage
generation scheme and, in particular, allow one  to smoothly tune
the period of the dynamical wiggler by rotating the diffraction
grating. There is a possibility of smooth frequency tuning
 for both the pump and the signal waves  by either
varying the geometric parameters of the volume diffraction grating
or by  rotating the  diffraction grating or the beam.  Moreover,
the VFEL allows one to create a dynamical wiggler in a large
volume, which is hard to achieve with  a static wiggler.

 There are two stages in the generation scheme proposed above (see Section 16, Fig. 16 \cite{LANL98}):

(a) creation of a dynamical wiggler in a system with
two-dimensional (three-dimensional) gratings (in other words,
during this stage the electromagnetic field, which exists inside
the VEFL resonator, is used to create a dynamical wiggler).
Smoothly varying the orientation of the diffraction grating in the
VEFL resonator, one can  smoothly change the dynamical wiggler
parameters;

(b) radiation is generated by the electron beam  interacting with
the dynamical wiggler, which has been created during the previous
stage (stage a).
\newline Both stages evolve in the same volume.

%%%%%%%%%%%%%%%%%%%%
\section{Conclusion}

Thus,
\begin{enumerate}
 \item
 volume FELs provide  frequency tuning by rotation of
the diffraction grating;
\item
 use of a non-one-dimensional DFB  reduces the generation
threshold and the size of the generation zone;
 \item VFELs allow
the use of a wide electron beam (or several beams) and diffraction
gratings of large volumes.
 Two- or three-dimensional diffraction gratings (artificial
crystals, now often called the electromagnetic or photonic
crystals) allow  distributing  the interaction over a  large
volume and overcoming the power restrictions on resonators and
thus open up the possibility of developing powerful generators
with wide electron beams (or system of beams);
\item VFELs enable
effective mode selection in oversized systems, where the radiation
wave length is significantly smaller than the resonator
dimensions;
  \item
  principles of VFEL can be used for creation of a
dynamical wiggler  with a variable period in a large volume;
\item
  two-stage scheme of generation, based on a non-one-dimensional distributed feedback,  can be applied for lasing
in the teraHertz frequency range using low-relativistic beams;
\item
  two-stage scheme of generation combined with the volume
distributed feedback can also form the basis for the  development
of  powerful generators with wide electron beams (or system of
beams); \item  use of a hybrid system composed of several
phase-locked vircator arrays generating modulated beams that enter
the VFEL resonator enables further increase of the generator's
power output \cite{LANL98,hybrid}. Moreover, a VFEL resonator
composed of periodically-arranged metallic elements aids in
preventing the Coulomb repulsion of electrons. The maximum
limiting current that can be transmitted through such a resonator
is greater than the limiting current that can be transmitted
through a vacuum drift space of a conventional BWO or TWT (compare
with the methods of increasing the current transmitted through the
resonator, suggested in \cite{R75,Gr73,rich}).

To achieve this, two(three)-dimensional electromagnetic crystals
formed by periodically arranged screens with periodically spaced
holes for beam transmission can also be used along with those
formed by periodically arranged wires, wire arrays and cylinders.
\end{enumerate}

Let us note in conclusion that inverse free electron lasers are
known to be used for particle acceleration. Similarly, VFELs with
a non-one-dimensional distributed feedback, arising in
electromagnetic (photonic) crystals, can be operated in the
inverse mode for accelerating particle beams  \cite{Vacuum}.  In
this case, the acceleration rate is appreciably higher than that
achieved in conventional FELs.

\end{document}